\documentclass[twocolumn]{IEEEtran}
\usepackage{amsfonts}
\usepackage[latin1]{inputenc}
\usepackage[spanish]{babel}
\newtheorem{theorem}{Theorem}
\newtheorem{acknowledgement}[theorem]{Acknowledgement}
\input tcilatex

\begin{document}

\author{Jos\'e Antonio Belinch\'on\thanks{E-mail: jabelinchon@alehop.com}\\
\emph{Grupo Inter-Universitario de An\'alisis Dimensional\\
Dept. Física ETS Arquitectura UPM\\
Av. Juan de Herrera 4 Madrid 28040\\
España}}
\title{Bulk Viscous Cosmological Model with G and Lambda Variables Through Dimensional Analysis}

\maketitle
\begin{abstract}
{\em A model with flat FRW symmetries and $G$ and $\Lambda ,$ variable is
considered in such a way that the momentum-energy tensor that describes the
model is characterized by a bulk viscosity parameter. For this tensor the
conservation principle is taken into account. In this paper it is showed how
to apply the dimensional method in order to solve the outlined equations}
{\bf in a trivial way}.
\end{abstract}
\begin{keywords}
Odes, AD, FRW Cosmologies, variable canstant
\end{keywords}
\section{\bf Introduction.}

Recently several models with FRW metric, where ``constants'' $G$ and $%
\Lambda $ are considered as dependent functions on time $t$ have been
studied. For these models, whose energy-momentum tensor describes a perfect
fluid, it was demonstrated that $G\propto t^\alpha $, where $\alpha $
represents a certain positive constant that depends on the state equation
imposed while $\Lambda \propto t^{-2}$ is independent of the state equation
(see \cite{A},\cite{T2}). More recently this type of model was generalized
by Arbab (see \cite{AB}) who considers a fluid with bulk viscosity (or
second viscosity in the nomenclature of Landau (see \cite{L})). The role
played by the viscosity and the consequent dissipative mechanism in
cosmology has been studied by many authors (see \cite{W}).\medskip).\

In the models studied by Arbab constants $G$ and $\Lambda $ are substituted
by scalar functions that depend on time $t$. The state equation that governs
the bulk viscosity is: $\xi \propto \xi _0\rho ^\gamma $ where $\gamma $ is
a certain indeterminate constant for the time being $\gamma \in \left[
0,1\right] $.\medskip

As we shall see, this problem is already solved, but our aim is to solve it
through Dimensional Analysis. We mean to point out how an adequate use of
this technique let us find the solution of the outlined equations in a
trivial way, even pointing out that it is not necessary to impose the
condition $div(T_{ij})=0.$ The paper is organized as follows: in the second
section the model is showed, expounding the equations and showing the
ingredients that compose the model. The third section is devoted to revise
the solution, which is reached by means of standard techniques of ODEs
integration (see in special \cite{SI}), while in the forth section the
dimensional technique will be developed in order to solve the model. This
section is divided in two subsections. In the first one, titled ``{\em %
pretty simple method'', }we point out how a naive use of D.A. brings us to
find such solution in a trivial way. Several cases are also studied here by
Arbab I. Arbab (see \cite{AB}), while in the other subsection a finer
dimensional technique is showed. That is why we call it ``{\em not so simple
method}''. This section is based on dimensional techniques (groups and
symmetries, see \cite{BIR}), in order to reduce the number of variables
intervening in the expounded ODEs. They are so simplified that its
integration is immediate. We think that the technique showed here is so
powerful that it shall be proved that imposing the condition $div(T_{ij})=0$
is not necessary to impose in order to solve the equations.

\section{\bf The model.}

This problem was posed by Arbab (see \cite{AB}). The equations of the model
are:

\begin{equation}
\label{e7}R_{ij}-\frac 12g_{ij}R-\Lambda (t)g_{ij}=\frac{8\pi G(t)}{c^4}%
T_{ij}
\end{equation}
and it is imposed that%
\footnote{we shall see that this condition it is not necessary to impose it}:%
$$
div(T_{ij})=0
$$
where $\Lambda (t)$ represent (stand) the cosmological ``constant''. The
basic ingredients of the model are:

\begin{enumerate}
\item  The line element defined by:
\begin{equation}
\label{e8}ds^2=-c^2dt^2+f^2(t)\left[ \frac{dr^2}{1-kr^2}+r^2\left( d\theta
^2+\sin {}^2\theta d\phi ^2\right) \right]
\end{equation}
we only consider here the case $k=0.$

\item  The momentum-energy tensor defined by:
$$
T_{ij}=(\rho +p^{*})u_iu_j-pg_{ij}
$$
where $\rho $ is the energy density and $p^{*}$ represents pressure $\left[
\rho \right] =\left[ p^{*}\right] $. The effect of viscosity is seen in:
\begin{equation}
\label{ee2}p^{*}=p-3\xi H
\end{equation}
where: $p$ is the thermostatic pressure, $H=\left( f^{\prime }/f\right) $
and $\xi $ is the viscosity coefficient that follows the law:
\begin{equation}
\label{ee3}\xi =k_\gamma \rho ^\gamma
\end{equation}
where $k_\gamma $ makes the equation be homogeneous i.e. it is a constant
with dimensions and where the constant $\gamma \in \left[ 0,1\right] $. And $%
p$ also verifies the next state equation:
\begin{equation}
\label{ee4}p=\omega \rho \qquad \omega =const.
\end{equation}
where $\omega \in \left[ 0,1\right] $ (i.e. it is a pure number) so that the
momentum-energy tensor verifies the so-called energy conditions.
\end{enumerate}

The field equations are:
\begin{equation}
\label{a1}2\frac{f\,^{\prime \prime }}{f\,}+\frac{(f\,^{\prime })^2}{f\,^2}%
=- \frac{8\pi G(t)}{c^2}p^{*}+c^2\Lambda (t)\ \
\end{equation}
\begin{equation}
\label{a2}3\frac{(f\,^{\prime })^2}{f\,^2}=\frac{8\pi G(t)}{\,c^2}\rho
+c^2\Lambda (t)\qquad \quad \
\end{equation}
deriving (\ref{a2}) and simplifying with (\ref{a1}) it yields
\begin{equation}
\label{SI1}\rho ^{\prime }+3(\omega +1)\rho H-9k_\gamma \rho ^\gamma H^2+
\frac{\Lambda ^{\prime }c^4}{8\pi G}+\rho \frac{G^{\prime }}G=0
\end{equation}
and at the moment we consider this other equation.
\begin{equation}
\label{a3}div(T_{ij})=0\text{ }\Leftrightarrow \rho ^{\prime }+3(\rho
+p^{*}) \frac{f^{\prime }}f=0
\end{equation}
if we develop the equation (\ref{a3}) we get:
\begin{equation}
\label{e9}\rho ^{\prime }+3(\omega +1)\rho H-9k_\gamma \rho ^\gamma H^2=0
\end{equation}

\section{\bf Non Dimensional Method.}

In this section we will mainly follow Singh et al work (see \cite{SI}). If
we take the equation (\ref{SI1}) regrouped, we get:
\begin{equation}
\label{m1}\stackunder{A1}{\underbrace{\rho ^{\prime }+3(\omega +1)\rho
H-9k_\gamma \rho ^\gamma H^2}}=\stackunder{A2}{\underbrace{-\left[ \rho
\frac{G^{\prime }}G+\frac{\Lambda ^{\prime }c^4}{8\pi G}\right] }}
\end{equation}
if take into account the conservation principle
\begin{equation}
\label{n5}\rho ^{\prime }+3(\omega +1)\rho H-9k_\gamma \rho ^\gamma H^2=0
\end{equation}
then we solve this equation by solving the equation $A2$ in (\ref{m1}), in
such a way that the equation to be solved is now:
\begin{equation}
\label{n6}\left[ \rho \frac{G^{\prime }}G+\frac{\Lambda ^{\prime }c^4}{8\pi G%
}\right] =0
\end{equation}
this equation is tried to be solved like this (see \cite{SI}). {\bf It is
defined} $\Lambda =\frac{3\beta H^2}{c^2}$ where $\beta $ is a numerical
constant, (hypothesis by Arbab (see \cite{AB}) as well as by Singh et al
(see \cite{SI}), condition that as we shall see, it is not necessary to
impose in the solution through D.A.) and from the equation (\ref{a2}) the
following relationship is obtained: $8\pi G\rho =3(1-\beta )H^2$. Hence if
all the equalities are replaced in the equation (\ref{n6}) it yields:
\begin{equation}
\label{n7}\frac 2{(1-\beta )}\frac{H^{\prime }}H=\frac{\rho ^{\prime }}\rho
\end{equation}
which is easily integrated.
\begin{equation}
\label{n8}H=C_1\rho ^{1/d}\qquad d=\frac 2{(1-\beta )}
\end{equation}
we get to the equation (\ref{n5}) with all these results
$$
\rho ^{\prime }+3(\omega +1)\rho H-9k_\gamma \rho ^\gamma H^2=0
$$
we arrive to the next equation:
\begin{equation}
\label{l1}\rho ^{\prime }+3C_1(\omega +1)\rho ^{\frac{d+1}d}-9C_1^2k_\gamma
\rho ^{\frac{d\gamma +2}d}=0
\end{equation}
which has got a particular solution in the case $\gamma =d^{-1}$ obtaining:
$$
\rho (t)=\dfrac 1{\left( a_0t\right) ^d}\qquad /\text{ }a_0=\left(
3C_1(\omega +1)-9k_\gamma C_1^2\right) d^{-1}
$$
and obtaining from it:
$$
f(t)=C_2t^{\dfrac 1{\left( 3(\omega +1)-3k_\gamma C_1\right) (1-\gamma )}}
$$
This is the most developed solution reached by Singh \ et al (see \cite{SI})
which is slightly different from the one by Arbab (see \cite{AB}).

\section{\bf Dimensional Method.}

We shall explore this section two dimensional methods. The first one,
probably the simplest one, has the inconvenience of having to depend on
Einstein criterion(see \cite{EIN} and Barenblatt \cite{B}), while the second
one is more powerful and more elaborated. We shall finish showing an
equation obtained without having to impose the condition $div(T_{ij})=0.$

\subsection{\bf Simple Method.}

The dimensional way followed in this section is probably the most basic and
simplest one. On one hand we integrate independently the equation
\begin{equation}
\label{e10}div(T_{ij})=0\text{ }\Leftrightarrow \rho ^{\prime }+3(\omega
+1)\rho \frac{f^{\prime }}f=0
\end{equation}
not taking into account the term $9k_\gamma \rho ^\gamma H^2,$ since if we
calculate its order of magnitude we verify that is very small $\approx
10^{-40}$ following, then, an asymptotic method (or perturbative) but this
must be justified from a physical and/or mathematical point of view$.$ If we
integrate the equation (\ref{e10}) it is obtained the well-known
relationship:
\begin{equation}
\label{e11}\rho =A_\omega f^{-3(\omega +1)}
\end{equation}
from this equation it is obtained one of the dimensional constants of our
problem: $A_\omega ,$ that has different dimensions and physical meaning
depending on the state equation imposed i.e. it depends on $\omega $. The
other dimensional constant considered has been obtained from the state
equation (\ref{ee3}) i.e. $\xi =k_\gamma \rho ^\gamma $, such constant $%
k_\gamma $ will also have different dimensions depending on the value $%
\gamma $, in such a way that the problem is reduced to the following set of
quantities and constants  $\frak{M}$.
$$
{\frak{M}} = (t,c,A_\omega ,k_\gamma ,a)
$$
where its respective dimensional equations in regard to a base $B=\left\{
L,M,T,\theta \right\} $ are (the base $B$ of this type of models has been
calculated in \cite{T}):
\begin{equation}
\label{e14p}
\begin{array}{c}
\left[ t\right] =T\quad \left[ c\right] =LT^{-1}\qquad \left[ a\right]
=L^{-1}MT^{-2}\theta ^{-4} \\
\left[ A_\omega \right] =L^{2+3\omega }MT^{-2}\qquad \left[ k_\gamma \right]
=L^{\gamma -1}M^{1-\gamma }T^{2n-1}
\end{array}
\end{equation}
where $a$ represents the radiation constant and it will be take into account
when we consider the thermodynamics quantities.\medskip\

Having done these considerations our aim is, therefore to solve this model
through D.A. The Pi-theorem will bring us to obtain two $\pi $ dimensionless
monomials; one of them will be the obtained in the case of a perfect fluid (%
\cite{T2}) and the other monomial will contain information on viscosity,
showing in this way that this type of models are very general, reproducing
the results obtained in the case for perfect fluids. Since all solutions
will depend on these two monomials we must take into account Barenblatt
criterion if we mean to reach a satisfactory final solution coincident with
the one obtained theoretically (see \cite{AB} and \cite{SI}).

\subsection{\bf Solutions through D.A.}

We shall calculate through D.A. i.e. by applying Pi-Theorem variation of $%
G(t)$ in function on $t$, energy density $\rho (t),$ the radius of Universe $%
f(t),$ the temperature $\theta (t)$, the entropy $S(t),$ the entropy density
$s(t)$ and finally the variation of the cosmological ``constant'' $\Lambda
(t).$ The dimensional method brings us to (see \cite{B} and \cite{T}):

\subsubsection{{\bf Calculation of }${\bf G(t):}$}

$G=G(t,c,A_\omega ,k_\gamma )$ where the dimensional equation of $G$
regarding to the base $B$ is: $\left[ G\right] =L^3M^{-1}T^{-2\text{ }}$.
Under this circumstances, the application of Pi-Theorem brings us to obtain
the following dimensionless monomials:%
$$
\left(
\begin{array}{rrrrrc}
& G & t & c & A_\omega & k_\gamma \\
L & 3 & 0 & 1 & 2+3\omega & \gamma -1 \\
M & -1 & 0 & 0 & 1 & 1-\gamma \\
T & -2 & 1 & -1 & -2 & 2\gamma -1
\end{array}
\right)
$$
$$
\pi _1=\frac{t^{1+3\omega }c^{5+3\omega }}{GA_\omega }\qquad \pi _2=\frac{%
ct^{1+\beta }}{A_\omega ^{(\gamma -1)\beta }k_\gamma ^\beta }
$$

It is observed that the first monomial ($\pi _1$) is identical to the one
obtained in the paper (\cite{T2}) for perfect fluids, while the second
monomial contains information on flow viscosity%
\footnote{these remarks, obviusly are valid for all the solutions obtained bellow}%
. These two monomials lead us to the following solution:
\begin{equation}
\label{r1}G\propto \frac{t^{1+3\omega }c^{5+3\omega }}{A_\omega }\cdot
\varphi \left( \frac{ct^{1+\beta }}{A_\omega ^{(\gamma -1)\beta }k_\gamma
^\beta }\right)
\end{equation}
where $\varphi $ represent an unknown function (i.e. at the moment we have
obtained a ``partial'' solution, in order to reach a more satisfactory
solution we must take into account the Barenblatt criterion) and $\beta $
is:
$$
\beta =\frac 1{3(\omega +1)(\gamma -1)}
$$

\subsubsection{{\bf Calculation of energy density }${\bf \rho (t)}$}

$\rho =\rho (t,c,A_\omega ,k_\gamma )$ regarding to the base $B,$ the
dimensional equation of the energy density is: $\left[ \rho \right]
=L^{-1}MT^{-2}$%
\begin{equation}
\label{r3}\rho \propto \frac{A_\omega }{\left( ct\right) ^{3(\omega +1)}}%
\cdot \varphi \left( \frac{ct^{1+\beta }}{A_\omega ^{(\gamma -1)\beta
}k_\gamma ^\beta }\right)
\end{equation}

\subsubsection{\bf Calculation of radius of Universe $f(t).$}

$f=f(t,c,A_\omega ,k_\gamma )$ where its dimensional equation is: $\left[
f\right] =L$%
\begin{equation}
\label{r4}f\propto ct\cdot \varphi \left( \frac{ct^{1+\beta }}{A_\omega
^{(\gamma -1)\beta }k_\gamma ^\beta }\right)
\end{equation}

\subsubsection{\bf Calculation of temperature $\theta (t).$}

$\theta =\theta (t,c,A_{\omega ,}a,k_\gamma )$ being its dimensional
equation: $\left[ \theta \right] =\theta $%
\begin{equation}
\label{r5}a^{\frac 14}\theta \propto \frac{A_\omega ^{\frac 14}}{\left(
ct\right) ^{\frac 34(1+\omega )}}\cdot \varphi \left( \frac{ct^{1+\beta }}{%
A_\omega ^{(\gamma -1)\beta }k_\gamma ^\beta }\right)
\end{equation}

\subsubsection{{\bf Calculation of entropy }$S(t).$}

$S=S(c,A_{\omega ,}a,k_\gamma ,t)$ where $\left[ S\right] =L^2MT^{-2}\theta
^{-1}.$
\begin{equation}
\label{r6}S\propto \left( A_\omega ^3a(tc)^{3(1-3\omega )}\right) ^{\frac
14}\cdot \varphi \left( \frac{ct^{1+\beta }}{A_\omega ^{(\gamma -1)\beta
}k_\gamma ^\beta }\right)
\end{equation}

\subsubsection{\bf Entropy density $s(t).$}

$s=s(t,c,A_{\omega ,}a,k_\gamma )$ where $\left[ S\right]
=L^{-1}MT^{-2}\theta ^{-1}$
\begin{equation}
\label{r7}s\propto \frac{\left( A_\omega ^3a\right) ^{\frac 14}}{\left(
ct\right) ^{\frac 94(1+\omega )}}\cdot \varphi \left( \frac{ct^{1+\beta }}{%
A_\omega ^{(\gamma -1)\beta }k_\gamma ^\beta }\right)
\end{equation}

\subsubsection{\bf Calculation of cosmological ``constant'' $\Lambda (t).$}

$\Lambda =\Lambda (t,c,A_\omega ,k_\gamma )$ being its dimensional equation $%
\left[ \Lambda \right] =L^{-2}$%
\begin{equation}
\label{r8}\Lambda \propto \frac 1{c^2t^2}\cdot \varphi \left( \frac{%
ct^{1+\beta }}{A_\omega ^{(\gamma -1)\beta }k_\gamma ^\beta }\right)
\end{equation}

\subsection{\bf Different Cases.}

All the following cases that we go on to study now have been studied by
Arbab (see \cite{AB}) confirming ``!`!'' his solution (\cite{SI}).\medskip\

In obtaining all solutions depending on two monomials we shall try to find a
solution to the problem expounded by means of the Barenblatt criterion (for
more details about the method used here see \cite{B} and \cite{T}).

\subsubsection{\bf $\gamma =1/2$ and $\omega =1/3,$ Radiation predominance.}

As we pointed out in the introduction the only models topologically
equivalent to the ones of classic FRW are those that follow the law $\xi
\propto \xi _0\rho ^{1/2}$ i.e $\gamma $$=1/2$ for its viscous parameter. In
this case we observe a Universe with radiation predominance $\omega =1/3.$
In order to obtain a complete solution we shall take into account Barenblatt
criterion since, we have obtained the solutions depending on an unknown
function $\varphi $. In this case the substitution of the values of $\omega $
and $\gamma $ leads us to:%
$$
G\propto \frac{t^2c^6}{A_\omega }\cdot \varphi \left( \frac{ct^{1/2}}{%
A_\omega ^{1/4}k_\gamma ^{-1/2}}\right)
$$

To get rid of the unknown function $\varphi $ we apply Barenblatt criterion,
for this purpose we need to know the order of magnitude of each monomial%
\footnote{see the table of numerical values at the end of the text}:%
$$
\pi _1=\frac{GA_\omega }{t^2c^6}\approx 10^{-10.59}\quad \pi _2=\frac{%
ct^{1/2}}{A_\omega ^{1/4}k_\gamma ^{-1/2}}\approx 10^{2.6}
$$
$$
\pi _1=\left( \pi _2\right) ^m\qquad m=\frac{\log \pi _1}{\log \pi _2}
$$
$$
G\propto \frac{t^2c^6}{A_\omega }\left( \frac{ct^{1/2}}{A_\omega
^{1/4}k_\gamma ^{-1/2}}\right) ^m\qquad /\ m\approx -4
$$
$$
G\propto \frac{c^2}{k_\gamma ^2}\qquad i.e.\qquad G\propto const.
$$
as we expected in having a model with $\gamma $$=1/2.$ We also obtain from
this point that $k_\gamma ^2=c^2/G.$ Whit regard to the rest of quantities
we operate identically finding without surprise that:%
$$
\rho \propto t^{-2}\quad f\propto t^{1/2}\quad \theta \propto t^{-1/2}\quad
S\propto t^0
$$
$$
s\propto t^{2/3}\quad \Lambda \propto const.
$$
As we see the model shows the same behavior in the principal quantities as
in the classic FRW model with radiation predominance.

Let see, for example, how $f$ has been calculated: Following the same steps
as we have seen in the case of calculations of $G$ it is observed that:
\begin{equation}
\label{r4}f\propto ct\cdot \varphi \left( \frac{ct^{1/2}k_\gamma ^{1/2}}{%
A_\omega ^{1/4}}\right)
\end{equation}
$$
\pi _1=\frac f{ct}\approx 10^{-2.6}\quad \pi _2=\frac{ct^{1/2}k_\gamma
^{1/2} }{A_\omega ^{1/4}}\approx 10^{2.6}
$$
$$
f\propto ct\left( \frac{ct^{1/2}k_\gamma ^{1/2}}{A_\omega ^{1/4}}\right)
^m\qquad /\quad m=-1
$$
$$
f\propto \left( \frac{cA_\omega ^{1/2}}{k_\gamma }\right) ^{\frac
12}t^{\frac 12}\propto \left( \frac{GA_\omega }{c^2}\right) ^{\frac
14}t^{\frac 12}
$$
This solution coincides with the one obtained for a classic FRW model with
radiation predominance. In any other cases $k_\gamma $ as well as $A_\omega $
will have other values to calculate.

\subsubsection{\bf $\gamma =1/2$ and $\omega =0.$ Matter predominance}

A model with matter predominance $\omega =0$ y topologically equivalent to a
classic FRW. In this case we find the following relationships:

Regarding to $G$ the solution obtained is (after replacing values $\gamma $
and $\omega )$:%
$$
G\propto \frac{tc^5}{A_\omega }\cdot \varphi \left( \frac{ct^{1/3}}{A_\omega
^{1/3}k_\gamma ^{-2/3}}\right)
$$
as we are working with a model described by matter instead of considering
energy density we find more convenient to consider matter density which
becomes a little dimensional readjustment in $A_\omega $ which becomes $%
\left[ A_\omega \right] =M$ in such a way that the solution pointed out
above for $G$ is still the following law:%
$$
G\propto \frac{tc^3}{A_\omega }\cdot \varphi \left( \frac{ctk_\gamma ^2}{%
A_\omega }\right) ^{1/3}
$$
as in the previous case we apply Barenblatt criterion which brings us to:%
$$
\pi _1=\frac{tc^3}{GA_\omega }=10^{-1.42}\qquad \pi _2=\left( \frac{%
ctk_\gamma ^2}{A_\omega }\right) ^{1/3}=10^{-0.47}
$$
$$
\pi _1=\left( \pi _2\right) ^m\qquad /m\approx -3
$$
$$
G\propto \frac{c^2}{k_\gamma ^2}\text{ \qquad i.e. \quad }G\propto const.
$$

In regard to the rest of quantities if we operate as before, we get:%
$$
\rho \propto \frac{c^2}{Gt^2}\quad f\propto \left( MG\right) ^{\frac
13}t^{2/3}\quad \Lambda \propto const.
$$
where we have used the equality $k_\gamma ^2=c^2/G$ and we have identified $%
A_\omega $ with the total mass of Universe $M$ i.e. The same behavior has
been obtained as in a FRW with matter predominance. Let see for instance how
we calculate radius $f:$

For this quantity the obtained solution is:%
$$
f\propto ct\cdot \varphi \left( \frac{ctk_\gamma ^2}{A_\omega }\right)
^{1/3}
$$
Barenblatt criterion brings us to:%
$$
\pi _1=\frac f{ct}=10^{0.5}\qquad \pi _2=\left( \frac{ctk_\gamma ^2}{%
A_\omega }\right) ^{1/3}=10^{-0.47}
$$
$$
\pi _1=\left( \pi _2\right) ^m\qquad /m\approx -1\qquad f\propto \left(
MG\right) ^{\frac 13}t^{2/3}
$$

\subsubsection{\bf $\gamma =3/4$ and $\omega =1/3.$ An Universe with
radiation predominance:}

In this case, as $\beta =-1$ we find the following solutions:%
$$
G\propto \frac{t^2c^6}{A_\omega }\cdot \varphi \left( \frac{ck_\gamma }{%
A_\omega ^{1/4}}\right)
$$
as the unknown function $\varphi $ does not depend on $t$ we can state
fearlessly that
$$
\varphi \left( \frac{ck_\gamma }{A_\omega ^{1/4}}\right) =D=const.
$$
since $c,$ $k_\gamma $ as well as $A_\omega $ are constant through
hypothesis, in such a way that%
$$
G\propto D^{\prime }t^2
$$
where $D^{\prime }=Dc^6/A_\omega .$ In this case we do not need to resort to
Barenblatt criterion in order to obtain a definitive solution. In regard to
the other quantities we obtain the following behaviors:%
$$
\rho \propto D\frac{A_\omega }{(ct)^4}\qquad f\propto Dct\qquad
a^{1/4}\theta \propto D\frac{A_\omega ^{1/4}}{ct}
$$
$$
S\propto D(A_\omega ^3a)^{1/4}\quad s\propto D\frac{(A_\omega ^3a)^{1/4}}{%
(ct)^3}\text{ \quad }\Lambda \propto D(ct)^{-2}
$$
In short, the obtained behaviors are:
$$
\begin{array}{c}
\rho \propto t^{-4}\quad \quad f\propto t\quad \quad \theta \propto t^{-1}
\\
S\propto const.\text{ \ }s\propto t^{-3}\quad \Lambda \propto t^{-2}
\end{array}
$$
this case follows an identical behavior to the one obtained in a model
described by a perfect fluid with $G$ and $\Lambda $ variables (see \cite{A}
and \cite{T2}) showing in this way the generality that we have obtained when
considering a bulk viscous fluid.

\subsubsection{\bf $\gamma =2/3$ and $\omega =0$ An Universe with matter
predominance.$.$}

In this case also $\beta =-1$, finding the following relationships as in the
previous case:%
$$
G\propto \frac{tc^5}{A_\omega }\cdot \varphi \left( \frac{ck_\gamma }{%
A_\omega ^{1/4}}\right)
$$
that in the previous case leads us to:
$$
G\propto D\frac{c^5t}{A_\omega }
$$
where $D=\varphi \left( \frac{ck_\gamma }{A_\omega ^{1/4}}\right) .$
Simplifying in the same way, without difficulty we reach:%
$$
\rho \propto t^{-3}\quad f\propto t\quad \Lambda \propto t^{-2}
$$
These two last cases are identical to the ones studied in references (see
\cite{A} and \cite{T2}).

\section{\ {\bf \ \ Not so simple method.}}

In this section we will combine dimensional techniques with standard
techniques of ODEs integration. With the dimensional method, we go on to
obtain dimensionless monomials, which will be replaced in the equations.
Thus, the number of variables will be reduced in such a way that the
resulting equation is integrable in a trivial way. We study two cases, the
first in which we consider {\bf $div(T_{ij})=0,$} while in the other, as we
shall see, such hypothesis is not needed.

\subsubsection{{\bf Considering the condition $div(T_{ij})=0.$ .}}

In this case we shall pay attention to the equation:
$$
\rho ^{\prime }+3(\omega +1)\rho H-9k_\gamma \rho ^\gamma H^2+\rho \frac{%
G^{\prime }}G+\frac{\Lambda ^{\prime }c^4}{8\pi G}=0
$$
taking into account the relationship $div(T_{ij})=0$ The following equality
is brought up:
$$
\stackunder{A1}{\underbrace{\rho ^{\prime }+3(\omega +1)\rho H-9k_\gamma
\rho ^\gamma H^2}}=\stackunder{A2}{\underbrace{-\left[ \rho \frac{G^{\prime }%
}G+\frac{\Lambda ^{\prime }c^4}{8\pi G}\right] }}
$$

The idea is the following: By using D.A. we obtain two $\pi -$monomials,
which are replaced in the equation, achieving a huge simplification of it.
On the other hand we integrate ($A1)$ and $(A2),$ solving completely in this
way the problem, this time without Barenblatt. let see. The monomials
obtained are: $\ \pi _1=\rho k_\gamma ^{\frac{-1}{1-\gamma }}t^{\frac{-1}{%
\gamma -1}}$ and $\pi _2=\Lambda c^2t^2$ \ i.e.
$$
\rho =ak_\gamma ^{\frac 1{1-\gamma }}t^{\frac 1{\gamma -1}}\qquad \Lambda
=\dfrac d{c^2t^2}
$$
where $a$ and $d$ are numerical constants. In a generic way the solution is
of the following form: $\rho =ak_\gamma ^{\frac 1{1-\gamma }}t^{\frac
1{\gamma -1}}$ \ if we define $b=\frac 1{1-\gamma }$ then $\rho =ak_\gamma
^bt^{-b}$ where $a=const.\in {\Bbb R}$ then $\rho ^{\prime }=-bak_\gamma
^bt^{-b-1}$ (paying attention only to the term $(A1)$ of the equation) it
yields:
\begin{equation}
-bak_\gamma ^bt^{-b-1}+3(\omega +1)ak_\gamma ^bt^{-b}H-9k_\gamma \left(
ak_\gamma ^bt^{-b}\right) ^\gamma H^2=0
\end{equation}
that simplifying it is reduced to:
\begin{equation}
9a^{\left( \gamma -1\right) }\left( f^{\prime }\right) ^2-3wt^{-1}ff^{\prime
}+bt^{-2}f^2=0
\end{equation}
\begin{equation}
f^{\prime }=\frac ft\left[ \frac 1{6a^{\gamma -1}}\left( w\pm
(w^2-4ba^{\gamma -1})^{\frac 12}\right) \right]
\end{equation}
where $w=(\omega +1),$ if it is defined
\begin{equation}
D=\left[ \frac 1{6a^{\gamma -1}}\left( w\pm (w^2-4ba^{\gamma -1})^{\frac
12}\right) \right]
\end{equation}
then, the solution has the following form:
\begin{equation}
f=lBt^D
\end{equation}
where $l$ is a certain numerical constant and $B$ is an integration constant
with dimensions, that can be identified with our result by making $%
B=A_\omega k_\gamma $.

Now we shall solve the other term of the equation (the $A2)$. the equation ($%
\left[ \rho \frac{G^{\prime }}G+\frac{\Lambda ^{\prime }c^4}{8\pi G}\right]
=0$ (\ref{n6})) can be solved in a trivial way if we follow the next
results. If we replace the monomials $\ \pi _1=\rho k_\gamma ^{\frac{-1}{%
1-\gamma }}t^{\frac{-1}{\gamma -1}}$ \ and $\pi _2=\Lambda c^2t^2$ \ in such
equation the integration of it becomes trivial:
$$
ak_\gamma ^{\frac 1{1-\gamma }}t^{\frac 1{\gamma -1}}\left( \frac{G^{\prime }%
}G\right) -\frac{dc^2}{4\pi Gt^3}=0
$$
\begin{equation}
G^{\prime }=\frac{dc^2}{a4\pi k_\gamma ^b}t^{b-3}\Longrightarrow G(t)=g\frac{%
dc^2}{a4\pi k_\gamma ^b}t^{b-2}
\end{equation}
where $a,d$ and $g\in {\Bbb R}$ (they are pure numbers). We can also observe
that this integral needs not be solved since a more careful analysis about
the number of $\pi -$monomials that we can obtain from the equation leads us
to obtain a solution of the type:%
$$
G=G(k_\gamma ,c,t)
$$
which brings us to:
$$
G(t)=gk_\gamma ^{-b}c^2t^{b-2}
$$
This method, as we have seen, is more elaborated and the solution,
therefore, finer though coincident with the previous one.

\subsubsection{{\bf Case in which $div(T_{ij})=0$ is not considered.}}

Let see now how we can tackle this problem from the D.A. point of view,
without imposing the condition $div(T_{ij})=0$. The base $B$ as before, is
still $B=\left\{ L,M,T\right\} $ while the fundamental set of quantities and
constants this time is $M=\left\{ t,c,k_\gamma \right\} $, with these data
we can obtain the following monomials from the equation\
\begin{equation}
\label{BER1}\rho ^{\prime }+3(\omega +1)\rho H-9k_\gamma \rho ^\gamma
H^2+\rho \frac{G^{\prime }}G+\frac{\Lambda ^{\prime }c^4}{8\pi G}=0
\end{equation}
considering that:
\begin{equation}
\label{BER2}\rho =ak_\gamma ^{\frac 1{1-\gamma }}t^{\frac 1{\gamma
-1}}\qquad \Lambda =\dfrac d{c^2t^2}
\end{equation}
these two monomials are replaced into the equation, which is quite
simplified:
$$
-bak_\gamma ^bt^{-b-1}+3(\omega +1)ak_\gamma ^bt^{-b}H-9k_\gamma \left(
ak_\gamma ^bt^{-b}\right) ^\gamma H^2+
$$
\begin{equation}
\label{BER3}+ak_\gamma ^bt^{-b}\frac{G^{\prime }}G-\frac{dc^2}{4\pi Gt^3}=0
\end{equation}
simplifying this equation, it yields:
\begin{equation}
\label{BER4}-9a^{\left( \gamma -1\right) }tH^2+3wH-bt^{-1}+\frac{G^{\prime }}%
G-\frac{dc^2}{4\pi ak_\gamma ^b}\dfrac{t^{b-3}}G=0
\end{equation}
that along with the field equations (\ref{a1}) and (\ref{a2}) carry us to
the next set of equations. For example we note that
$$
3H^2=a\frac{8\pi G}{c^2}k_\gamma ^bt^{-b}+\frac d{t^2}
$$
that we replace into the equation that we are treating, resulting:
$$
-bt^{-1}+3w\left( \frac{a8\pi k_\gamma ^b}{3c^2}Gt^{-b}+\frac d{3t^2}\right)
^{\frac 12}-
$$
$$
-9a^{\left( \gamma -1\right) }\left( \frac{a8\pi k_\gamma ^b}{3c^2}%
Gt^{-b}+\frac d{3t^2}\right) t+\frac{G^{\prime }}G-\frac{dc^2}{4\pi
ak_\gamma ^b}\dfrac{t^{b-3}}G=0
$$
that solving it results:
\begin{equation}
G=gk_\gamma ^{-b}c^2t^{b-2}
\end{equation}
where $g\in {\Bbb R}$ represents a numerical constant. We finally observe
that as in the previous section we could have taken into account the three
monomials obtained from the equation i.e.
$$
\rho =ak_\gamma ^bt^{-b}\qquad \Lambda =\dfrac d{c^2t^2}\qquad G=g\frac{%
c^2t^{b-2}}{k_\gamma ^b}
$$
replacing them into the equation%
$$
\rho ^{\prime }+3(\omega +1)\rho H-9k_\gamma \rho ^\gamma H^2+\rho \frac{%
G^{\prime }}G+\frac{\Lambda ^{\prime }c^4}{8\pi G}=0
$$
and calculate $f,$ arriving at the same solution obtained in the above
section i.e.%
$$
f=lBt^D
$$
We have proved that it is not necessary to impose the condition $%
div(T_{ij})=0$ since it is obtained, in this case, the same solution as the
one obtain imposing it.\

\section{\bf Conclusions.}

We have studied a cosmological model described by a momentum-energy tensor
characterized by a fluid with bulk viscosity, in which, furthermore, we have
considered the constants $G$ and $\Lambda $ as functions depending on time
i.e. as variables and we have imposed the condition $div(T_{ij})=0$. We have
proved how a suitable use of Dimensional Analysis enables us to find the
solution of such model in a ``{\bf trivial'' }way. With the ``{\em Pretty
simple method}'', we have obtained two $\pi -$monomials, one of them is the
one obtained in the case for a perfect fluid (\cite{T2}) and the other
monomial contains the information about viscosity, showing, in this way,
that this type of models is very general being able to reproduce the result
obtained in the case of a perfect fluid. In order to solve the problem we
have taken into account Barenblatt criterion being able to arrive to obtain
a complete solution of the problem. Standing out that our results coincide
with the solutions obtained by Arbab I. Arbab \cite{AB}. We have shown too
that with the ``{\em not so simple method}'' we arrive to solve the problem
without necessity of impose any condition. We believe, nevertheless, that
the ``{\em simple method}'' can be more effective, since, we obtain more
solutions with it or more complete solutions in the sense of finding in it
solutions such as $\Lambda \propto t^{-2}$ as well as $\Lambda =const.$
while the ``{\em not so simple method}'' the only solution that is obtained
is $\Lambda \propto t^{-2}$, but has the drawback of depending on Barenblatt
criterion i.e. we depend on the always insecure numerical data.\newpage\ \

\begin{acknowledgement}
I wish to thank to Javier Aceves for helping with the translation into English
\end{acknowledgement}

\begin{table}
  \begin{tabular}{|l|l|l|l|l|l|l|l|l|} \hline
    {\bf}& {\bf $G$ }  &  {\bf $c$ }  &  {\bf$\rho $  } &{\bf$f$ } &{\bf$A_\omega $} &{\bf$k_\gamma $} & {\bf$\theta $} & {\bf$t$}  \\ \hline
        $\omega =1/3$ & -10.17& 8.47 & -13.379 & 26 & 90.62 & 13.5  & 0.436 & 20.25  \\ \hline

        $\omega =0$ & -10.17 & 8.47   & -26.397 & 26 & 54     & 13.5   &         & 17  \\ \hline
  \end{tabular}
  \caption{The values refer to a logarithmic scale i.e. $G\approx 10^{-10.17}$ etc... meassured in the International System $\left\{ m,kg,s\right\} $. In the case $\omega =0,$ $\rho $ represents mass density while in the case $\omega =1/3$ represents energy density.\label{key}}
\end{table}


\begin{thebibliography}{10}
\bibitem[1]{A}  {\bf A-M. M. Abdel-Rahman.} Gen. Rel. Grav. {\bf 22,} 655,
(1990). {\bf M. S. Bermann}. Gen. Rel. Grav. {\bf 23}, 465, (1991). {\bf %
Abdussaltar and R. G. Vishwakarma.} Class. Quan. Grav. {\bf 14}, 945, (1997).

\bibitem[2]{T2}  {\bf Belinch\'on, J.A}. (Physics/9811017)

\bibitem[3]{AB}  {\bf Arbab I. Arbab.}Gen. Rel. Grav. {\bf 29}, 61-74,
(1997).

\bibitem[4]{L}  {\bf Landau, L.D. and Lifshitz, E. M}. Fluid Mechanics
(Pergamon, London 1976).\

\bibitem[5]{W}  {\bf Weinberg,S.} Gravitation and cosmology. (Wiley, N.Y.
1972) pp. 593-594. {\bf Murphy, G.L.} Phys. Rev. D12, 4231, (1973). {\bf %
Padmanabhan, T, Chitre, S.M.} Phys. Lett. A 120, 433, (1987).{\bf Barrow, J.
D.} Nuclear .Phys {\bf B310}, 743. (1988).

\bibitem[6]{B}  {\bf Barenblatt. }Scaling, self-similarity and intermediate
asymptotics. Cambridge texts in applied mathematics N 14 1996 CUP. {\bf %
Palacios, J. }Dimensional Analysis. Macmillan 1964 London. {\bf R. Kurth. }%
Dimensional Analysis and Group Theory in Astrophysic. Pergamon 1972.

\bibitem[7]{EIN}  {\bf Einstein, A}. Ann Phys. {\bf 35}, 679-694,(1911)

\bibitem[8]{BIR}  {\bf Birkhoff, G.} Hydrodynamics. Princeton U.P. 1960.
{\bf Cari\~nena, J. F. et at} Adv. Electr. Elec. Phys. {\bf 72}, 181, (1988)

\bibitem[9]{T}  {\bf Belinch\'on, J.A}. (Physics/9811016).

\bibitem[10]{SI}  {\bf T. Singh, A. Beesham, W.S. Mbokazi,} Gen. Rel. Grav.
{\bf 30}, 573, (1998).
\end{thebibliography}
\end{document}